\documentclass[twocolumn,amsmath,amssymb,superscriptaddress,longbibliography,pra]{revtex4-1}

\usepackage{color}
\usepackage{graphicx}
\usepackage{amssymb}
\usepackage{amsmath}
\usepackage{mathrsfs}
\newcommand{\ud}{\mathrm{d}}

\let\Eps\varepsilon
\usepackage[breaklinks=true]{hyperref}
\graphicspath{ {./figures/} {./}}

\begin{document}

\title{Viscous damping of  {gravity-capillary waves: dispersion relations and nonlinear corrections}}

\author{Andrea Armaroli}
\email{andrea.armaroli@unige.ch}
\affiliation{GAP-Nonlinearity\&Climate, Institute for Environmental Sciences, Universit{\'e} de Gen{\`e}ve, Boulevard Carl-Vogt 66, 1211 Gen{\`e}ve 4, Switzerland}

\author{Debbie Eeltink}
\affiliation{GAP-Nonlinearity\&Climate, Institute for Environmental Sciences, Universit{\'e} de Gen{\`e}ve, Boulevard Carl-Vogt 66, 1211 Gen{\`e}ve 4, Switzerland}

\author{Maura Brunetti}
\affiliation{GAP-Nonlinearity\&Climate, Institute for Environmental Sciences, Universit{\'e} de Gen{\`e}ve, Boulevard Carl-Vogt 66, 1211 Gen{\`e}ve 4, Switzerland}

\author{J{\'e}r{\^o}me Kasparian}
\affiliation{GAP-Nonlinearity\&Climate, Institute for Environmental Sciences, Universit{\'e} de Gen{\`e}ve, Boulevard Carl-Vogt 66, 1211 Gen{\`e}ve 4, Switzerland}

\begin{abstract}
	 {We discuss the impact of viscosity on nonlinear propagation of  surface waves at the interface of air and a fluid of large depth.} After a survey of the available approximations of the dispersion relation, we propose to modify the hydrodynamic boundary conditions to model both short and long waves. From them we derive a nonlinear Schr{\"o}dinger equation where both linear and nonlinear parts are modified by dissipation and show that the former plays the main role in both gravity and capillary-gravity waves while, in most situations, the latter represents only small corrections. This provides a justification of the conventional approaches to damped propagation found in the literature.  
\end{abstract}

\maketitle

\section{Introduction}

Deep-water surface waves are among the most studied examples of nonlinear physical systems. The nonlinearity stems from kinematic and dynamic boundary conditions at the air-fluid interface. Triplet \cite{mcgoldrick1965} and quadruplet \cite{Zakharov1968} interactions can be included and a general integro-differential equation  derived.  A suitable expansion of the integral kernel \cite{Trulsen2000,Stiassnie1984}  was shown to connect it to simple propagation models, like the nonlinear Schr{\"o}dinger equation (NLS) \cite{SulemNLSBOok} and its generalizations (modified NLS---MNLS), the best known among which is the Dysthe equation \cite{Dysthe1979}. The universal NLS possesses many remarkable properties and solutions, such as the modulation instability (MI, also known as Benjamin-Feir instability---BFI) \cite{Benjamin1967,Lo1985,Zakharov2009}, solitons and breathers, the clearest examples of the balance of dispersion and nonlinearity.

In the propagation of deep-water waves, gravity effects dominate for long waves (small wave-numbers) and surface tension for short waves. Moreover, viscosity is the ubiquitous damping mechanism \cite{Lamb1945hydrodynamics,Landauvol6} (other routes to dissipation include impurities, obstacles, wave-breaking\dots)  {and plays a role in fundamental studies of analogue gravity \cite{Rousseaux2018}}.
%
%
It induces a vorticity in the fluid and requires therefore to solve the full Navier-Stokes (NS) equations. For  small viscosity, the vorticity is significant only in a small boundary layer close to the surface \cite{Lamb1945hydrodynamics,Landauvol6,Longuet-Higgins1953,Longuet-Higgins1960}. 

For small amplitude waves, i.e.~when nonlinearities are negligible, the velocity potential formulation valid for an inviscid flow can be corrected to include the effects of viscosity \cite{Ruvinsky1991,Joseph2004,Wang2006}. Different loss rates and corrections to the group velocity can be found in Ref.~\cite{Padrino2007}. The availability of a quasi-potential formulation greatly simplifies the numerical solution of the hydrodynamic problem \cite{West1987,Dommermuth1987}.  {Moreover, it is often assumed without justification that the linear dissipation is so small that no nonlinear correction is required for the formulation of a modified NLS including damping \cite{Wu2006,Dias2008}. Alternatively, nonlinear damping mechanisms were proposed in the form of a Landau damping \cite{Fabrikant1980} or in a phenomenological way \cite{Kato1995,Schober2015a}. }

In this work we  quantify the nonlinear corrections to propagation equations stemming from kinematic viscosity and justify therefore the conventional assumption of Refs.~\cite{Wu2006,Dias2008}. Our approach is based on the dissipation method  detailed in \cite{Prosperetti1976,  Padrino2007}, but can be generalized to any expression for the dispersion relations and  damping mechanisms. After recalling some simple considerations about dispersion relations and justifing the choice of a specific form (Section \ref{sec:VDR}), we propose a modified set of hydrodynamic equations consistent with our approach (section \ref{sec:HyEq}). These are straightforwardly adapted in  Section \ref{sec:sNLS} to generalize the 1D MNLS \cite{Trulsen2000} 
\begin{equation}
	\frac{\partial \hat B}{\partial t} + i\left(\omega(k_0+\kappa)- \omega_0\right)\hat B + i\frac{\omega_0 k_0^2}{2} \mathcal{F}_\kappa\left[|B|^2B\right]=0,
\label{eq:MNLS}
\end{equation}  
with   {$\hat{B}(\kappa,t)\equiv\mathcal{F}_\kappa[B(x,t)]=\frac{1}{\sqrt{2\pi}}\int\ud x B(x,t)e^{-i\kappa x}$ the Fourier transform from real space  of $x$ (in the co-moving frame) to the relative wavenumber space of $\kappa$. In fact, the envelope $B(x,t)$ of the surface elevation is centered around the carrier wavenumber $k_0$; $\omega(k)$ the real-valued dispersion relation ($\omega_0\equiv\omega(k_0)$). }
We quantify the nonlinear corrections to the recurrence period and spectral-mean downshift to be less than one percent in typical experimental conditions of a water tank. 
We also verify that the full dispersion relation plays a key role in the nonlinear evolution of the MI, e.g., in explaining the frequency downshift first observed in \cite{Lake1977}. 
Finally (section \ref{sec:tNLS}), we show that the considered dispersion relation can be inverted so that not only the space-like, but also the time-like formulation of the MNLS naturally generalizes to the dissipative case. 
The same approach can be adapted to high-order NLS, such as the Dysthe \cite{Dysthe1979} or the recently proposed compact \cite{Dyachenko2011} and super-compact \cite{Dyachenko2017} equations. 
Conclusions and outlook complete our manuscript.

\section{Viscous dispersion relations}
\label{sec:VDR}
We first briefly review how to express the dispersion relation for deep-water  capillary-gravity waves at the surface of an incompressible viscous fluid. The solution of the linearized hydrodynamic equations for 1D propagation in the $x$-direction is a plane wave $\exp(i k x - i \omega(k)t)$.
We decompose $\omega(k)=\omega_\mathrm{R}(k) + i \omega_\mathrm{I}(k)$, where $\omega_\mathrm{R}>0$ (resp.~$\omega_\mathrm{R}<0$) represents forward- (resp.~backward-) propagating waves  {(for $k>0$, otherwise  the opposite applies)} and $\omega_\mathrm{I}<0$ is the damping rate. 

 {In the limit of infinite depth,} the dispersion relation can be shown to be the solution $\omega(k)$ of the implicit equation  \cite{Lamb1945hydrodynamics,Landauvol6} 
\begin{equation}
\left(2-i\frac{\omega}{\nu k^2}\right)^2 + \frac{|k|(g + sk^2)}{\nu^2 k ^4} = 4\left(1-i\frac{\omega}{\nu k^2}\right)^{\frac{1}{2}}
\label{eq:disprelimp}
\end{equation}
where $g$ is the standard acceleration due to gravity, $s \equiv T/\rho_\mathrm{f}$ with $T$ the surface tension (in $\mathrm{N m^{-1}}$) of the fluid-air interface and $\rho_\mathrm{f}$ is the density of the fluid. The gravity and capillary contributions dominate, respectively, for small and large $k$,  i.e.~for large and small wavelength. Finally, $\nu$ denotes the kinematic viscosity of the fluid (in $\mathrm{m^2/s}$) and is the physical origin of  $\omega_\mathrm{I}$.

The detailed derivation of Eq.~\eqref{eq:disprelimp} from the linearized Euler equations for an incompressible fluid, along with their kinematic and boundary conditions,  consists in including a vorticity field and assuming that the mass transport occurs only in a boundary layer close to the air-fluid interface \cite{Lamb1945hydrodynamics,Landauvol6,Longuet-Higgins1953, Longuet-Higgins1960, Ruvinsky1991, Longuet-Higgins1992a,Dias2008}. Eq.~\eqref{eq:disprelimp} being quartic, two forward-traveling wave branches exist: one is unphysical (it corresponds to a velocity potential diverging for infinite depth $z\to-\infty$, see also in section \ref{sec:HyEq}).
\begin{figure}[hbtp]
\centering
\includegraphics[width=.45\textwidth]{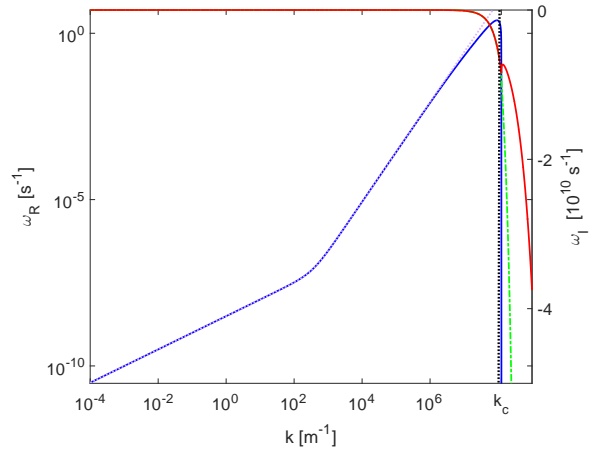}
\caption{Dispersion relation for gravity-capillary waves propagating at the interface of air and water in the presence of kinematic viscosity. Numerical solution of Eq.~\eqref{eq:disprelimp} for forward propagating waves ($\omega_R\ge 0$).  {On the left axis (in logarithmic scale), the solid blue and dotted lines refer to $\mathrm{Re}\,{\omega}$ and its  inviscid  counterpart $\tilde\omega$. The red dotted and green dashed-dotted lines pertain to the right axis (linear scale) and show the two branches of $\mathrm{Im}\,{\omega}$.   The vertical dotted line marks to the cut-off $k_c$ beyond which $\omega_\mathrm{R}=0$ and $\omega_\mathrm{I}$ bifurcates in two standing wave branches, yielding different damping rates.} }
\label{fig:waterdisp}
\end{figure}
The physical solution is shown in Fig.~\ref{fig:waterdisp} for a gravity-capillary wave propagating at the interface of air and water ($s = 7.28\times 10^{-5}\mathrm{\,m^3\,s^{-2}}$) in the presence of kinematic viscosity $\nu=1\times 10^{-6}\mathrm{\,m^2\,s^{-1}}$.   {For large $k$, in the  $\mathrm{\mu m}^{-1}$ range}, $\omega_\mathrm{R}$ rises and then drops abruptly. At $k=k_c=1.25\times 10^{8}\,\mathrm{m}^{-1}$ it cuts off, i.e.~the viscous damping is so strong that $\omega_\mathrm{R} = 0$: the mode is a standing wave. For $k>k_c$ two branches with different damping are admitted, see the right axis of Fig.~\ref{fig:waterdisp}. The least-damped branch (solid line) is expected to take the leading physical role at those wave-numbers,  {the other (dashed line) being dissipated much more rapidly \cite{Landauvol6} and not behaving as  a surface wave \citep{Mainardi1987}.}

Relying on physical arguments\cite{Ruvinsky1991,Joseph2003,Joseph2004,Wang2006,Padrino2007}, it was shown that simpler explicit formulas can be obtained, by expressing the vorticity in  terms of the velocity potential and surface elevation by assuming linear propagation and small viscosity.
Nevertheless, a simple algebraic manipulation of Eq.~\eqref{eq:disprelimp} allows us to re-derive them easily. Let us define  $\theta\equiv -i\frac{\omega}{\nu k^2}$ and $\tilde \theta\equiv -i\frac{\tilde\omega}{\nu k^2}$ (related to the Reynolds number defined for water waves), with $\tilde\omega(k)\equiv\sqrt{|k|(g+sk^2)}$ (i.e.~the inviscid dispersion relation). 
Eq.~\eqref{eq:disprelimp} is written compactly as 
\begin{equation}
\left(\frac{2+\theta}{\tilde \theta}\right)^2 -  1 = \frac{4}{\tilde \theta^{\frac{3}{2}}}\left(\frac{1+\theta}{\tilde \theta}\right)^\frac{1}{2}
\label{eq:diseprelimp1}
\end{equation}

 {Notice that for $\nu k^2\ll 1$, $\tilde{\theta}\gg 1$. Moreover, for small $k$ it can be safely assumed that the quantity between parentheses in the LHS is of order one, i.e.~$\theta\approx\tilde\theta$, while the RHS can be considered of higher order, by virtue of its prefactor.} 
The standard dispersion relation is indeed obtained by neglecting the RHS of Eq.~\eqref{eq:diseprelimp1}: we write $\theta = \pm\tilde \theta-2$, i.e.~
\begin{equation}
	\omega = \pm\tilde\omega -2i\nu k^2,
	\label{eq:stddisp}
\end{equation}
which we refer to as (small-$k$) Lamb approximation \cite{Lamb1945hydrodynamics}. It was shown that its physical justification can be traced back to the smallness of the vortical contribution to pressure and the fact that the  surface deformation and boundary layer (where the vorticity is non-negligible) are small \cite{Ruvinsky1991,Dias2008}. However, the viscosity enters only in the imaginary part, so no cut-off for traveling waves can possibly appear. Lamb derived also the two branches of the damping rate in the opposite case of small $\tilde{\theta}$, which read respectively $\omega_\mathrm{I} = -\frac{\tilde\omega^2}{2\nu k^2}$ and $\omega_\mathrm{I} = -0.91 \nu k^2$, the former being the most physically important. The cut-off is estimated from Eq.~\eqref{eq:diseprelimp1}, by looking for a real solution $\theta$ of double multiplicity.  {We solve the system composed by Eq.~\eqref{eq:diseprelimp1} and its derivative with respect to $\theta$. We obtain $\theta=\beta$, with $\beta$ the real root of $\beta^3 +5\beta^2+8\beta+3=0$. Thus, $\theta$ is eliminated and $k_c$ is the solution of $\tilde\omega^2(k_c)=(4/\alpha-\alpha^2)(\nu k_c^2)^2\approx 0.58 (\nu k_c^2)^2$, where  $\alpha\equiv\beta+2$.}

In order to have a single expression for small and large $k$, different approximations of the RHS of Eq.~\eqref{eq:disprelimp} are shown to behave better than Eq.~\eqref{eq:stddisp}.
Indeed, if we let $1+\theta \approx 1$ on the RHS of Eq.~\eqref{eq:diseprelimp1}, we re-obtain the result of the dissipation method (DM), \cite{Prosperetti1976,Padrino2007}, i.e.~
\begin{equation}
\theta=-2\pm\sqrt{\tilde \theta^2+4}.
\label{eq:DM}
\end{equation}
In contrast to Eq.~\eqref{eq:stddisp}, this relation exhibits a  cut-off in the real part of $\omega$, for $\tilde{\theta}=-2i$. Beyond, the two signs represent the two branches of dissipation of the standing mode. 

Further, we notice that, in  the alternative approach proposed in \cite{Joseph2004} (viscous potential flow---VPF), only the dynamic boundary condition (Bernoulli equation) is modified, by introducing viscosity as an external pressure perturbation, analogously to \cite{Wu2006}. It can be reproduced by the expansion $(1+\theta)^\frac{1}{2} \approx 1 + \theta/2$ and its solution is written, in our notation
\begin{equation}
\theta =-1\pm\sqrt{\tilde \theta^2+1}.
\label{eq:VPF}
\end{equation} 
Finally, we may ask ourselves what is the result of expanding the RHS of Eq.~\eqref{eq:diseprelimp1} to the second order: we write a third approximation, 
\begin{equation}
\theta=-\frac{2}{3}\pm\frac{2}{3}\sqrt{ \frac{3}{2}\tilde\theta^2 + 1},
\label{eq:MVPF}
\end{equation} 
which we will refer to as modified VPF (MVPF). 

 {The results are summarized in Table \ref{tab:formulas}. 
Among the  dispersion relations \eqref{eq:DM}--\eqref{eq:MVPF}, only DM corresponds asymptotically to Lamb solution as far as damping at small $k$ is concerned; $k_\mathrm{c}$ is not accurate, though.} The real part of the VPF is a better approximation at the cut-off and matches very well with the conventional dispersion relation, but the damping is half of the expected one at small $k$, while standing waves ($k>k_c$) of lesser damping have the correct asymptotic  form (compared to the estimation made by Lamb, see above). Finally the MVPF behaves better at the cut-off, but does not reproduce the behavior of either $\omega_\mathrm{R}$  or $\omega_\mathrm{I}$ for small values of $k$, so it is of little practical use in the most accessible oceanic regimes. In \cite{Wang2006}, it was shown that irrotational theories fail to provide a good approximation around $k_c$. The MVPF shows instead that a good approximation of the cut-off is incompatible to a satisfactory asymptotic behavior at small and large $k$ simultaneously. 

\begin{table}
\centering
\begin{tabular}{|r|c|c|c|c|}
\hline
& $\omega_\mathrm{R}(k\ll 1)$ & $\omega_\mathrm{I}(k\ll 1)$ & cut-off &  $\omega_\mathrm{I}(k\gg 1)$\\
\hline
Implicit \eqref{eq:diseprelimp1} & $\tilde \omega$ & $-2\nu k^2$ & $\tilde\omega^2=0.58(\nu k^2)^2$
& \vtop{\hbox{\strut$-\frac{\tilde\omega^2}{2\nu k^2}$}\hbox{\strut$-{0.91\nu k^2}$}
}\\
\hline
DM \eqref{eq:DM}& $\tilde \omega$ & $-2\nu k^2$ & $\tilde\omega^2=4(\nu k^2)^2$ 
&\vtop{\hbox{\strut $-\frac{\tilde\omega^2}{4\nu k^2}$}\hbox{\strut $-4\nu k^2$}}\\
\hline
VPF \eqref{eq:VPF}& $\tilde \omega$ & $-\nu k^2$ & $\tilde\omega^2=(\nu k^2)^2$ 
& \vtop{\hbox{\strut $-\frac{\tilde\omega^2}{2\nu k^2}$}\hbox{\strut $-2\nu k^2$} }\\
\hline
MVPF \eqref{eq:MVPF}& $\sqrt{\frac{3}{2}}\tilde \omega$ & $-\frac{2\nu k^2}{3}$ & $\tilde\omega^2=\frac{2}{3}(\nu k^2)^2$ & \vtop{\hbox{\strut  $-\frac{\tilde\omega^2}{2\nu k^2}$}\hbox{\strut $-\frac{4}{3}\nu k^2$}}\\
\hline
SDM \eqref{eq:SDM}& $\tilde \omega$ & $-2\nu k^2$ & n/a & $-\frac{\tilde\omega^2}{2\nu k^2}$\\
\hline
\end{tabular}
\caption{Comparison of the asymptotic behavior of the implicit and approximated dispersion relations. The classical estimations of Lamb on the implicit formula, the DM, VPF, and MVPF are simply derived in the text from the full dispersion relation, SDM is the simplified DM found by neglecting the viscous terms under the square-root in Eq.~\eqref{eq:denominators}. The first two are also discussed in \cite{Padrino2007}. }
\label{tab:formulas}
\end{table}

\begin{figure}
\centering
\includegraphics[width=.45\textwidth]{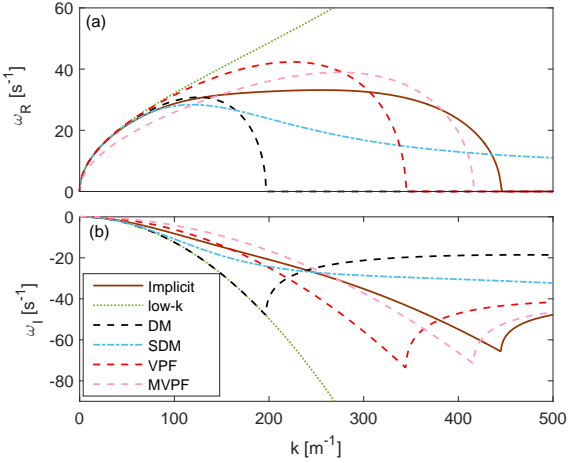}
\caption{Dispersion relation for gravity-capillary waves propagating at the interface of air and glycerin. (a) real part; (b) imaginary part. Numerical solutions of Eq.~\eqref{eq:disprelimp} for forward propagating waves ($\omega_R\ge 0$) are shown as a solid dark yellow line. The conventional approximation is shown by a green dotted line. The different approximations are represented by dashed lines (black---DM, red---VPF, pink---MVPF). Finally, the dashed-dotted lines correspond to the SDM: it does not exhibit a clear cut-off, but provides a good approximation of damping at both $k$ limits.	}
\label{fig:glycerindisp}
\end{figure}

The numerical values of cut-off for the fluids considered in \cite{Padrino2007} are reported in Table \ref{tab:cutoff}, where we include also the MVPF results, for the sake of completeness.
\begin{table}[h]
\begin{tabular}{|r|c|c|c|}
\hline
& Water & Glycerin & SO10000\\
\hline
Implicit \eqref{eq:diseprelimp1}& $1.25\times 10^8$ &  445.18 & 54.30 \\
\hline
DM \eqref{eq:DM} & $1.82\times 10^7$ & 196.81 & 28.50\\
\hline
VPF \eqref{eq:VPF}& $7.28\times 10^7$ & 344.64  & 45.29 \\
\hline
MVPF \eqref{eq:MVPF}& ${1.09\times 10^8}$ &  416.10 &  51.86\\
\hline
\end{tabular}
\label{tab:cutoff}
\caption{Cut-off values (in $\mathrm{m^{-1}}$) for the three examples of Ref.~\cite{Padrino2007} and their comparison to the MVPF result.  {SO10000 stands for Silicone oil of viscosity 10000 cSt, for which in SI units $\rho_\mathrm{f}=9.69 \times 10^2 \,\mathrm{kg\, m^{-3}}$, $\nu=1.02\times 10^{-2}\mathrm{\,m^2\,s^{-1}}$, $s = 2.10\times 10^{-5}\mathrm{\,m^3\,s^{-2}}$)} }
\end{table}
In order to visually confirm the formulas of table \ref{tab:formulas}, we compare the different approximations in Fig.~\ref{fig:glycerindisp} for a surface between glycerin and air ($\rho_\mathrm{f}=1.26 \times 10^3 \, \mathrm{kg\, m^{-3}}$, $\nu=6.21\times 10^{-4}\mathrm{\,m^2\,s^{-1}}$, $s = 5.03\times 10^{-5}\mathrm{\,m^3\,s^{-2}}$). The larger viscosity allows us to have a cut-off for relatively small $k$ and observe both short and long wave ranges in linear scale. The same behavior applies to other fluids.

We include also a further simplification. Suppose we choose the DM relation of Eq.~\eqref{eq:DM}, which  provides the best approximation of damping and group velocity for long waves. We rewrite it as \begin{equation}
\omega(k) = \frac{\tilde\omega}{\hat{D}(k)}
\label{eq:DM1}
\end{equation}
with
\begin{equation}
\hat{D}(k)\equiv {\pm\sqrt{1-\left(\frac{2\nu k^2}{\tilde{\omega}}\right)^2} + \frac{2i\nu k^2}{\tilde{ \omega}}}. 
\label{eq:denominators}
\end{equation}
For small  $\nu k^2$, we can neglect the term proportional to $\nu^2$ under the square root in Eq.~\eqref{eq:denominators} and obtain the simplified DM (SDM)
\begin{equation}
\omega\approx  \frac{\tilde\omega}{\pm 1 + \frac{2i\nu k^2}{\tilde{ \omega}}},
\label{eq:SDM}
\end{equation}
It is trivial to verify---see also table \ref{tab:formulas}---that Eq.~\eqref{eq:SDM} provides a better asymptotic behavior than Eq.~\eqref{eq:DM} for damping at large $k$, at the expense of a smooth transition around the cut-off, as shown in Fig.~\ref{fig:glycerindisp} as a cyan dash-dotted line.
That is, the series expansion of $\tilde\omega/\omega$ provides a more robust approximation than the conventional direct Taylor expansion of $\omega$ (i.e.~the low-$k$ Lamb approach, green dotted lines in Fig.~\ref{fig:glycerindisp}), which predicts ever increasing phase velocity $v_\mathrm{p}\equiv\omega_\mathrm{R}/k$ and damping. This is  similar to the fit of a dispersion relation by means of a Pad{\'e} approximant, well known in optics \cite{Amiranashvili2010a}. 
%


\section{Hydrodynamic equations}
\label{sec:HyEq}
Viscosity induces vorticity, thus making the solution of the nonlinear hydrodynamic problem (NS equations) extremely complicated. The most practical workaround is to  extend the use of a velocity potential $\phi(x,z,t)$, where $z$ is the depth coordinate and $x$ is the longitudinal propagation direction, to the viscous case. 

By denoting the free surface elevation $\eta(x,t)$, the system of hydrodynamic equations in an inviscid and infinitely deep fluid reads as
\begin{equation}
\begin{aligned}
\phi_{xx} + \phi_{zz} &= 0 &\mathrm{for}\; -\infty<z<\eta \\
\nabla \phi &\to 0 &\mathrm{for}\; z\to-\infty	\\
\eta_t + \phi_x \eta_x   - \phi_z & =0  &\mathrm{for}\; z=\eta	\\
\phi_t + \frac{1}{2}\left(\phi_x^2 + \phi_z^2\right) + g \eta &\\
-s \frac{\eta_{xx}}{\left(1+\eta_x^2\right)^{\frac{3}{2}}} & = 0 &\mathrm{for}\; z=\eta	\\
\end{aligned}
\label{eq:HydroInviscid}
\end{equation}
These are respectively the Laplace equation in the fluid, the rigid bottom condition, the kinematic and the dynamic boundary conditions at the free surface.
In the linear limit, the solutions of Eq.~\eqref{eq:HydroInviscid} are plane waves, $\eta =  \eta_0\exp(ikx- i\omega t)$ and $\phi =  \phi_0\exp(ikx- i\omega t + |k|z)$.  The inviscid dispersion relation $\omega = \tilde\omega(k)$ is the compatibility condition of the homogeneous system
\begin{equation}
\begin{bmatrix}
i\omega & |k|\\
-(g+sk^2) & i\omega
\end{bmatrix}
\begin{bmatrix}
\eta_0 \\ \phi_0
\end{bmatrix} = 0.
\label{eq:invhomog}
\end{equation}
 
Eq.~\eqref{eq:stddisp} is obtained by the substitution $i\omega\to i\omega -2\nu k^2$ in Eq.~\eqref{eq:invhomog}. In the wavenumber domain, we obtain a correction $-2\nu k^2 \eta_0$ to the kinematic boundary condition and $-2\nu k^2 \phi_0$ to the dynamic one. 
To transform them back to the spatial domain, we notice that $\eta$ does not depend on $z$ and $\phi$ is the solution of the Laplace equation. The operator correspondence $ik\leftrightarrow \frac{\partial}{\partial x}$ (established by the plane-wave definition) is correct for both terms and allows us to obtain the well-known weakly viscous hydrodynamic system \cite{Ruvinsky1991,Dias2008}
\begin{equation}
\begin{aligned}
\phi_{xx} + \phi_{zz} &= 0 &\mathrm{for}\; -\infty<z<\eta \\
\nabla \phi &\to 0 &\mathrm{for}\; z\to-\infty	\\
\eta_t + \phi_x \eta_x  - \phi_z &= 2\nu \eta_{xx} &\mathrm{for}\; z=\eta	\\
\phi_t + \frac{1}{2}\left(\phi_x^2 + \phi_z^2\right) + g \eta &\\
-s \frac{\eta_{xx}}{\left(1+\eta_x^2\right)^{\frac{3}{2}}}  = 2\nu \phi_{xx} & = -2\nu \phi_{zz} &\mathrm{for}\; z=\eta	\\
\end{aligned}
\label{eq:HydroDDZ}
\end{equation}
Its physical motivation is that the velocity field can be  decomposed in a potential and a vorticity contribution, $(u,w)=(\phi_x-\Omega^y_z,\phi_z+\Omega^y_x)$.  The vorticity pseudo-vector has only the $y$-component, $\Omega\equiv (0,\Omega^y,0)$, and is expressed  as a function of $\phi$ and $\eta$ by using the linearized boundary conditions and by assuming $\Omega^y_z\approx 0$ \cite{Ruvinsky1991,Dias2008}. 
 {Strictly speaking, neglecting this term violates the conservation of mass. This will be discussed in details in a future publication.}
The fully nonlinear NS equations couple different velocity components and is hard to write in as simple terms as Eqs.~\eqref{eq:HydroInviscid} and \eqref{eq:HydroDDZ}.

We propose a solution to this difficulty based on the approximation presented in the previous section. 
The principle behind our choice is that, in Eq.~\eqref{eq:HydroInviscid} as well as in the unidirectional models derived from it (e.g.~the NLS or the Dysthe equation \cite{Dysthe1979}), the energy is the sum of a kinetic part, which depends on dispersion, and a potential part, which is ascribed to nonlinear interaction. A periodic exchange between the two parts characterizes, e.g., the nonlinear stage of MI. Both are damped by kinematic viscosity is thus plausible. Since the Hamiltonian density which encompasses the different energy terms is associated to the evolution in time, the associated operator must be modified as a whole.
This allows us to better quantify the role of dissipation in the non-linear propagation of waves. 
Consider Eq.~\eqref{eq:DM1}: it is the solution of 
\[
\begin{vmatrix}
i \omega \hat D(k) & |k|\\
-(g+sk^2) & i\omega \hat{D}(k)
\end{vmatrix} = 0.
\]
Thus, in Eq.~\eqref{eq:HydroInviscid}, we make the substitution $\frac{\partial}{\partial t}\to \bar\partial_t\equiv \mathcal{D}(-i\frac{\partial}{\partial x})\frac{\partial}{\partial t}$, where  $\mathcal{D}(-i\frac{\partial}{\partial x})$ is  the operator in the physical space associated in the Fourier space to $\hat{D}(k)$ in Eq.~\eqref{eq:denominators}.

This allows us to formally obtain an alternative form for the hydrodynamic equations
\begin{equation}
\begin{aligned}
\phi_{xx} + \phi_{zz} &= 0 &\mathrm{for}\; -\infty<z<\eta \\
\nabla \phi &\to 0 &\mathrm{for}\; z\to-\infty	\\
\bar\partial_t\eta + \phi_x \eta_x   - \phi_z & =0  &\mathrm{for}\; z=\eta	\\
\bar\partial_t \phi+ \frac{1}{2}\left(\phi_x^2 + \phi_z^2\right) + g \eta &\\
-s \frac{\eta_{xx}}{\left(1+\eta_x^2\right)^{\frac{3}{2}}} & = 0 &\mathrm{for}\; z=\eta	\\
\end{aligned}
\label{eq:HydroNew}
\end{equation}

The proposed system reduces to Eq.~\eqref{eq:HydroDDZ} for $\nu k^2\ll \tilde\omega$, by replacing $-1/i\tilde\omega\leftrightarrow \int\ud t$ and $-k^2\leftrightarrow \frac{\partial^2}{\partial x^2}$. 


 Eq.~\eqref{eq:HydroNew}  provides however an alternative set of equations to be employed in full hydrodynamic solvers, like the high-order spectral method (HOSM) \cite{West1987,Dommermuth1987,Touboul2010,Kharif2010}. 

 {We stress that the choice of DM or SDM is not binding. A similar approach could be used with the other approximations discussed in Sec.~\ref{sec:VDR}, the numerical solution of Eq.~\eqref{eq:disprelimp}, as well as a different fit based on experimental data.}

In the next section we will discuss how to derive a nonlinear NLS-like propagation equation and show that small albeit measurable differences between the  linear and nonlinear dissipation terms exist.

\section{A space-like propagation equation}
\label{sec:sNLS}

In order to generalize the MNLS of Eq.~\eqref{eq:MNLS} to the dissipative case, we can follow the  approach of \cite{Trulsen2000}, based on  Zakharov's method \cite{Zakharov1968}, to provide the justification for an NLS in which a general dispersion relation and not a polynomial truncation thereof is included. 
The free-surface elevation is reconstructed from the envelope $B(x,t)$ as $\eta= \frac{1}{2}\left[b(x,t)+\mathrm{c.c.}\right]$, where $b \equiv B e^{ -i(k_0 x -\omega_0 t) }$  is the free-wave component and c.c.~denotes the complex conjugate. Notice that the plane wave factors are defined to be  real: $\omega_0\equiv\omega_\mathrm{R}(k_0)$.  


Eq.~\eqref{eq:MNLS}, where  the nonlinearity is assumed small and viscosity is neglected ($\omega_R=\tilde\omega$), is written for $\hat{b}(k,t)\equiv\mathcal{F}_k[b(x,t)]$ as
\begin{equation}
	\hat{b}_t +i\tilde\omega(k)\hat b +i\frac{\omega_0 k_0^2}{2}\mathcal{F}_k[|b|^2 b]  =0
\end{equation}
from which, by means of the substitution $\partial_t \to \bar\partial_t$, we derive the equation of motion for 
\begin{equation}
	\frac{\partial \hat b}{\partial t} + i\frac{\tilde\omega(k)}{\hat D(k)}\hat b + i\frac{\omega_0 k_0^2}{2\hat D(k)} \mathcal{F}_k\left[|b|^2b\right]=0.  
\label{eq:VNP}
\end{equation}
Back to the slowly-varying variable $B$, we can write, for $\hat B(\kappa,t)=\mathcal{F}_\kappa[B(x,t)]$
\begin{equation}
	\frac{\partial \hat B}{\partial t} + i\left[\frac{\tilde\omega(k_0+\kappa)}{\hat D(k_0  + \kappa)}-\omega_0\right]\hat B + i\frac{\omega_0 k_0^2}{2\hat D(k_0 + \kappa)} \mathcal{F}_\kappa\left[|B|^2B\right]=0.  
\label{eq:VNLS}
\end{equation}
This represents our generalization of Eq.~\eqref{eq:MNLS} in the dissipative case. 


By Taylor-expanding it to fourth order in $\kappa$ and writing the resulting terms in the physical space by replacing powers of $\kappa$ by derivatives in $x$, we obtain the linear terms of Ref.~\cite{Carter2016}.  This approach may prove convenient also for generalizing a forced MNLS model \cite{Eeltink2017}.

\begin{figure}[hbtp]
\centering
\includegraphics[width=0.45\textwidth]{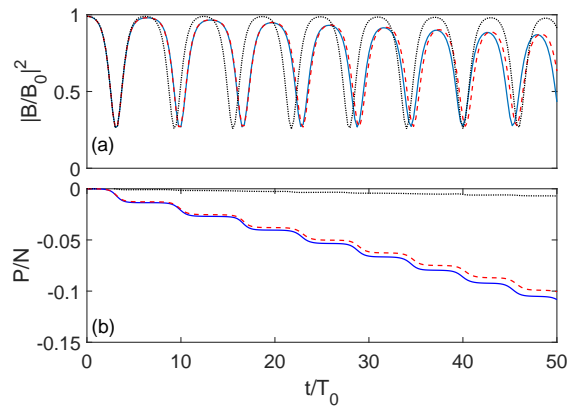}
\caption{Comparison of the nonlinear evolution of MI with linear and nonlinear damping. Time is normalized by nonlinear time $T_0=(2\Eps^2k_0)^{-1}$. $k_0=10$, $\varepsilon=0.1$, $\nu=1\times 10^{-6}$, $\alpha = 0.01$, surface tension is neglected. (a) Oscillations of the central peak (Stokes wave); (b) spectral mean $\kappa_\mathrm{m}$. The solid blue lines represent the full model (Eq.~\eqref{eq:VNLS}), the dashed red (resp.~black dotted) lines are obtained by neglecting $\hat D(k)$ in the denominators of the nonlinear (resp.~linear) part.}
\label{fig:NLSMIcomp}
\end{figure}

\begin{figure}[t]
\centering
\includegraphics[width=0.45\textwidth]{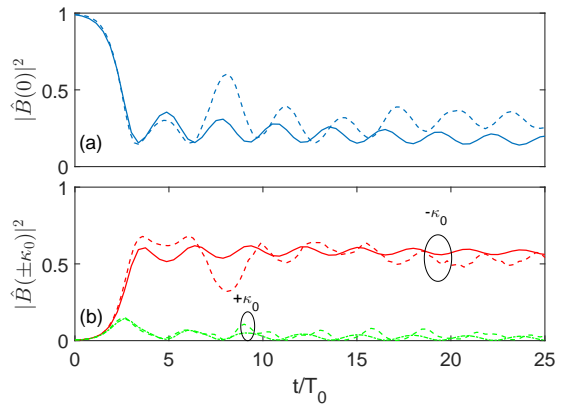}
\caption{Comparison of results with purely linear (solid lines) and purely nonlinear (dashed lines) damping. Parameters are $k_0=75$, $\varepsilon=0.1$, $\nu=1\times 10^{-6}$, $s=7.28\times 10^{-5}$, $\alpha = 0.01$. (a) Central mode $\kappa=0$; (b) main unstable modes $\kappa=\pm\kappa_0$. We notice that the downshifted mode at $-\kappa_0$ soon acquires most of the energy. While the linear damping irreversibly stops recurrence, the nonlinear damping is not sufficient. }
\label{fig:NLSdownshift}
\end{figure}

Below, in Figs.~\ref{fig:NLSMIcomp}-\ref{fig:NLSdownshift}, we show that the dispersive contribution explains alone most of the frequency downshift observed in experiments. Our approach provides also a nonlinear viscous damping (the imaginary part of the nonlinearity),  
valid for small $k$ where the bound modes (small corrections to $\eta$ oscillating at integer multiples of $k_0$ and enslaved, for pure gravity waves, to the free-modes $B$) are not resonantly excited.
It also introduces a wavenumber-dependent correction to nonlinearity (proportional to $\nu^2$ under the square root in $\hat{D}$, see Eq.~\eqref{eq:DM1}). 
We expand the nonlinear damping coefficients as
$
\frac{2 \nu k^2}{\omega} \frac{\omega_0 k_0^2}{2}  \approx \nu k_0^3\left[ k_0+ \left(2- \frac{v_\mathrm{g}^0}{v_\mathrm{p}^0}\right) \kappa\right]
$,
where $v_\mathrm{p}^0\equiv\tilde\omega(k_0)/k_0$ and $v_\mathrm{g}^0\equiv \tilde\omega'(k_0)$ are the phase and group velocities at $k_0$, respectively, neglecting viscosity.
The first contribution is a homogeneous nonlinear damping (which can be obtained independently by the method of multiple scales  \cite{Eeltink2018}), while the second is a derivative damping term, i.e., the dissipative counterpart of the self-steepening in nonlinear optics \cite{Agrawal2012}. Its effect is small, because energy dissipation caused by linear attenuation limits the bandwidth of $B$.

As an example, we show in Fig.~\ref{fig:NLSMIcomp} the solution for a harmonically perturbed Stokes wave propagating at the water-air interface ($B(x,0)=B_0\left[\sqrt{1-\alpha}+\sqrt{2\alpha}\cos{\kappa_0 x}\right]$), with initial steepness $\varepsilon\equiv \frac{B_0 k_0}{\sqrt{2}}=0.1$, $k_0 = 10$, $\alpha=1\times 10^{-2}$ and neglecting surface tension. The perturbation wave-number is $\kappa_0 = 2\varepsilon k_0\sqrt{2B_0}$, i.e.~the maximally unstable mode predicted by the NLS. We notice that the linear and nonlinear dissipation scale as $2\nu k_0^2$ and $2\nu k_0^4|B|^2\approx 2\nu k_0^2 \varepsilon^2$, respectively. We thus expect  that only at extreme steepness and breather peaks is the nonlinear damping  non-negligible with respect to the linear one. 

In Fig.~\ref{fig:NLSMIcomp} we compare the dynamics of Eq.~\eqref{eq:VNLS} (full model), with the results of neglecting $\hat{D}(k)$ in either the linear or nonlinear part (we refer to them as nonlinear and linear damping, respectively). The comparison shows both the energy attenuation and downshift of the spectral mean $\kappa_{\rm m}\equiv P/N$, with
\begin{equation}
N= \int\ud x |B|^2,\, P= \mathrm{Im}\,\int\ud x B_xB^*,
\label{eq:NPspace}
\end{equation} 
respectively, the norm and the momentum of the field. 
$\kappa_\mathrm{m}$ depends mainly on the linear damping, see App.~\ref{app:downshift}. As shown in \cite{Armaroli2018}, the slightest amount of dissipation causes the recurrence period to stretch and a period-1 orbit to be attracted to a period-2 one. 

This behavior is found in results of the full model (blue solid lines) and in the linearly damped (dashed red lines) simulations. The discrepancy in recurrence periods is just a contraction of about 1$\%$ (per period). In panel (a), we notice also that the nonlinear damping alone (black dotted line) leads instead to a behavior more similar to the undamped result. As far  as the downshift of  $\kappa_\mathrm{m}$ is concerned [panel (b)], the full model exhibits about  1$\%$ more shift than the linearly damped one. Notice that the difference between the two nearly equals  the pure nonlinear contribution (black dotted line). 

The blue solid lines exhibits the same behaviour if  Eq.~\eqref{eq:SDM} is used instead of Eq.~\eqref{eq:DM1} (not shown). The discrepancies between the solid and dashed lines, observed in Fig.~\ref{fig:NLSMIcomp}, are mainly explained by the nonlinear damping $\nu k_0^4 |B|^2 B$: viscous corrections to group velocity have an even smaller impact on nonlinear coefficients.


This behavior does not qualitatively change if we include  {Dysthe nonlinear terms} in Eq.~\eqref{eq:VNLS}, as in, e.g., \cite{Stiassnie1984}: the downshift is associated to a transient upshift at each recurrence cycle \cite{Armaroli2018}. 

A different dynamics is observed for shorter wavelengths even though stronger damping limits the impact of nonlinear phenomena.
We thus simulated the same initial conditions as above, with $k_0=75$, i.~e.~near the transition from gravity to capillary waves. We show in Fig.~\ref{fig:NLSdownshift}  the results of the purely linear and purely nonlinear dissipation models. In panel (a) we plot the central mode ($\kappa=0$) and in panel (b) the two main unstable modes at each side of it ($\kappa=\pm\kappa_0$).  Even during the initial MI phase the energy is converted into a pair of unstable sidebands in an asymmetric fashion, which favors the lower wave-number. As soon as most of the energy is located at $-\kappa_0$ (at  $t/T_0\approx2.5$), we observe a remarkable difference: the central mode cannot recover, not even partially, its initial condition and stabilizes below 0.5 for pure linear damping (solid lines), while a regime of more erratic recurrence, in which every 5-6 cycles the energy is transferred back to the central mode (above 0.6), is observed for pure nonlinear damping (dashed lines).   {The full model (with linear and nonlinear damping, not shown) exhibits almost the same behaviour as the purely linear case.} This is consistent with the interplay of surface tension and viscosity to favor a permanent downshift of the spectral peak, observed in several full numerical simulations of the Euler system \cite{Skandrani1996,Dias1999}. 


We finally remark that the present nonlinear viscous damping does not exclude the existence of other nonlinear loss mecahnisms. We would like to mention the  Landau damping \cite{Fabrikant1980}, which plays a role at large $k$, where the bound modes are resonantly excited and dissipated by viscous damping. Its even symmetry prevents however a downshift of either the spectral  peak or mean. Alternatively, a loss mechanism like the  $\beta$-term introduced in  \cite{Kato1995,Schober2015a} explains the frequency downshift observed in the nonlinear stage of MI, by virtue of its odd symmetry but has no clear physical origin: it may represent a model of wave-breaking, not included here. Nevertheless, we have shown that the spectral mean shifts permanently simply due to the variation of damping with frequency. 

Both examples, Figs.~\ref{fig:NLSMIcomp} and \ref{fig:NLSdownshift}, show that an MNLS where the full dispersion (with dissipation) is included in the linear part is sufficient to explain at least qualitatively the partial recurrence and spectral downshift in the nonlinear behavior of MI. 
Thus the implicit assumption of \cite{Dias2008} that a nonlinear correction due to viscosity plays a minor role in deep-water wave propagation is confirmed in its physical soundness. 
 {In App. \ref{app:downshift}, we derive the expressions for the rate of change of the spectral mean and motivate more rigorously the downshifting trend. }

\section{A time-like propagation equation}
\label{sec:tNLS}
Laboratory conditions usually imply the measurement of the temporal profiles at different positions along a wave-tank by means of wave-gauges. The time-like formulation of the NLS and its generalizations are thus more practical than their  space-like counterparts for describing and interpreting measurements.

In order to obtain such a formulation, we need to invert the dispersion relation and derive an explicit expression for $k(\omega)$. The method we used above to derive the DM approximation provides a straightforward solution, at least for $s=0$, that reads
\begin{equation}
k(\omega)=\frac{\tilde k}{\hat D_k(\omega)}
\label{eq:invdisp}
\end{equation}
with $\tilde k (\omega) \equiv \frac{\omega^2}{g}$, the conventional dispersion of deep-water gravity waves, and $\hat D_k(\omega)\equiv\sqrt{1+16i\frac{\nu\omega^3}{g^2}}$ the correction due to viscosity, which contributes to both the real and imaginary parts of $k(\omega)$. Eq.~\eqref{eq:invdisp} can be simplified as $k(\omega)\approx \frac{\omega^2}{g} \left[1-4i\frac{\nu\omega^3}{g^2}\right]^{-1}$.

 As shown in Sec.~\ref{sec:sNLS}, the nonlinear damping is negligible in most cases. 
We write the dissipative time-like MNLS in the frequency domain---with respect to the detuning $\Omega\equiv\omega-\omega_0$ (here $\hat{B}(x,\Omega)\equiv\mathcal{F}_\Omega[B(x,t)]$)---as
\begin{equation}
	\frac{\partial \hat B}{\partial x} - i\left[\frac{\tilde  k(\omega_0+\Omega)}{\hat D_k(\omega_0 + \Omega)}-k_0\right]\hat B - i\frac{k_0^3}{2} \mathcal{F}_\Omega\left[|B|^2B\right]=0.  
\label{eq:tVNLS}
\end{equation}
This model can be applied to assess the effects of the  dispersive damping in a wave-tank experiment.

\section{Conclusions}
\label{sec:Concl}
After having recalled the different forms of dispersion relation for deep-water  gravity-capillary waves in the presence of viscosity and unified their derivation, we discussed their physical validity all over the wavenumber/frequency range. We exploited these results to reformulate the hydrodynamic equations and quantify the impact of kinematic viscosity on nonlinear damping. We showed that the simplest NLS model with full dispersion (in both the real and imaginary part) provides most of the justification for the downshift of the spectral mean during the nonlinear stage of evolution of the MI: corrections in the nonlinear behavior have only a small effect.  This provides an \emph{a posteriori} justification  of the choice of using  dispersion relations only in the linear part of a nonlinear propagation equation \cite{Dias2008,Trulsen2000}. 
 {However, this does not forbid consideration of other damping mechanisms, e.g.~wave-breaking.} Experiments are needed to determine the best form for a rational dispersion relation, which could be used even in fully nonlinear hydrodynamic simulations.

\begin{acknowledgments}
We acknowledge the financial support from the Swiss National
Science Foundation (Projects Nos.~200021-155970 and 200020-175697). We would like to thank John D.~Carter for fruitful discussions. 
\end{acknowledgments}

%

\appendix
\section{Downshift of the spectral mean}
\label{app:downshift}
In this appendix, we will derive from Eq.~\eqref{eq:VNLS} the rate of change of $N$ and $P$ defined above in the main text, Eq.~\eqref{eq:NPspace}, and of the spectral mean $P/N$.

We consider only the linear term of Eq.~\eqref{eq:VNLS},
\begin{equation}
\frac{\partial \hat B}{\partial t} + i\left[\frac{\tilde\omega(k_0+\kappa)}{\hat D(k_0  + \kappa)}-\omega_0\right]\hat B=0.
\label{eq:linVNLS}
\end{equation}
First, we rewrite Eq.~\eqref{eq:NPspace} as
\begin{equation}
N = \frac{1}{2\pi} \int\ud \kappa |\hat B|^2, 
\,P= \frac{1}{2\pi}\int \ud \kappa\, \kappa |\hat B|^2,
\label{eq:NPspectral}
\end{equation}
the fist by Parseval theorem and the second by simple manipulations of Fourier transforms. 

From Eqs.~\eqref{eq:linVNLS} and \eqref{eq:NPspectral}, by elementary substitution, we derive
\begin{equation}
\frac{\ud N}{\ud t} = \chi \int\ud \kappa (k_0+\kappa)^2 |\hat B|^2, \, \frac{\ud P}{\ud t}  = \chi\int\ud \kappa    (k_0+\kappa)^2 \kappa |\hat B|^2, 
\label{eq:NPvariation}
\end{equation}
with $\chi=-\frac{2\nu}{\pi}$.

Now, expanding each integral of Eq.~\eqref{eq:NPvariation}, we write 
\begin{equation}
\frac{\ud \kappa_\mathrm{m}}{\ud t}  =-\frac{4\nu}{N^2}\left[
2k_0(QN-P^2)+\left(SN-PQ\right)
\right],
\label{eq:kmvariaiton}
\end{equation}
where
\begin{equation}
Q = \frac{1}{2\pi} \int\ud \kappa \, \kappa^2 |\hat B|^2 =\int\ud x  |B_x|^2, 
\end{equation}
\begin{equation}
S= \frac{1}{2\pi}\int \ud \kappa\, \kappa^3 |\hat B|^2 =  \frac{1}{2i}\int\ud x  \left[B_x^*B_{xx}-\mathrm{c.c.}\right].
\end{equation}	
Notice that $Q$ corresponds to the same quantity defined in Ref.~\cite{Carter2016}.

We find that $QN-P^2\ge 0$ directly from Cauchy-Schwarz inequality, $|\langle \hat f,\hat g \rangle| \le \lVert f\rVert \lVert g\rVert$ with the two square-integrable functions $\hat f(\kappa)\equiv \hat B(\kappa)$ and $\hat g(\kappa)\equiv \kappa^{\frac{1}{2}}\hat B(\kappa)$. 

\begin{figure}[hbtp]
\centering
\includegraphics[width=.45\textwidth]{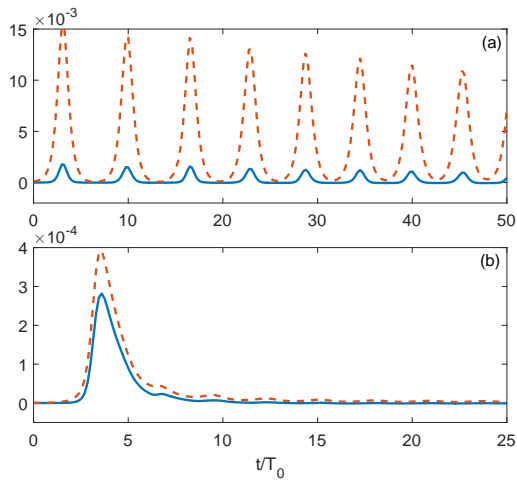}
\caption{Evolution in time of $SN-PQ$ (blue solid line) and $2k_0(QN-P^2)+\left(SN-PQ\right)$ (red dashed line) for surface waves at the air-water interface. Panel (a) [resp.~(b)] corresponds to the gravity (resp.~gravity-capillary) case of Fig.~\ref{fig:NLSMIcomp} [resp.~Fig.~ \ref{fig:NLSdownshift}] in the main text. While $SN-PQ$ is not always rigorously positive, so it is in most of the situations and the net mean-frequency shift, as predicted by Eq.~\eqref{eq:kmvariaiton} is always positive.}
\label{fig:downshift}
\end{figure}

 {
As far as $SN-PQ$ is concerned, nothing more can be stated mathematically. Numerical simulations show that its sign is positive during most of the evolution of surface waves at the air-water interface. This is displayed as solid blue lines in Fig.~\ref{fig:downshift}, where we consider the examples of Figs.~\ref{fig:NLSMIcomp}--\ref{fig:NLSdownshift}. Moreover, the overall sign of the terms in square brackets of Eq.~\eqref{eq:kmvariaiton} is positive in both examples (red dashed lines in Fig.~\ref{fig:downshift}), thus, according to Eq.~\eqref{eq:kmvariaiton}, the mean frequency shifts downwards, for both gravity and gravity-capillary regimes. These results hold also for larger values of steepness up to $\varepsilon\approx 0.3$, for which the use of NLS-like nonlinearity becomes questionable, and for smaller $k_0$ (now shown).
}

As far as the nonlinear terms are concerned, no rigorous explicit relation can be derived, but numerical simulations hint at their contribution to \emph{increase} the said downshift. 

\end{document}